\documentclass[conference]{IEEEtran}
\IEEEoverridecommandlockouts

\usepackage{cite}
\usepackage{amsmath,amssymb,amsfonts}
\usepackage{algorithmic}
\usepackage{graphicx}
\usepackage{textcomp}
\usepackage{xcolor}
\usepackage{siunitx}
\usepackage{multirow}

\def\BibTeX{{\rm B\kern-.05em{\sc i\kern-.025em b}\kern-.08emT\kern-.1667em\lower.7ex\hbox{E}\kern-.125emX}}    

\begin{document}

\title{Reconfigurable Slotted Antenna Inspired by Multidimensional Modulation\\
\thanks{This work was sponsored by the National Natural Science Foundation of China (NSFC) Grant Nos. 61805097 and 61935015.}
}

\author{\IEEEauthorblockN{Nan-Shu Wu}
\IEEEauthorblockA{\textit{College of Electronic Science and Engineering} \\
\textit{Jilin University}\\
Changchun, China \\
wuns18@mails.jlu.edu.cn}
\and
\IEEEauthorblockN{Su Xu\IEEEauthorrefmark{1}}
\thanks{\IEEEauthorrefmark{1} Su Xu and Hong-Bo Sun are corresponding authors.}
\IEEEauthorblockA{\textit{College of Electronic Science and Engineering} \\
\textit{Jilin University}\\
Changchun, China \\
xusu@jlu.edu.cn}
\and
\IEEEauthorblockN{Zuojia Wang}
\IEEEauthorblockA{
\textit{School of Information Science and Engineering} \\
\textit{Zhejiang University}\\
Hangzhou, China\\
z.wang@sdu.edu.cn}
\and
\IEEEauthorblockN{Hong-Bo Sun\IEEEauthorrefmark{1}}
\IEEEauthorblockA{
\textit{Department of Precision Instrument} \\
\textit{Tsinghua University}\\
Beijing, China \\
hbsun@tsinghua.edu.cn}
}

\maketitle

\begin{abstract}
Multidimensional modulation was widely studied in the past decades due to the explosive development of modern wireless communication. Here, we propose a spirally reconfigurable slotted antenna inspired by the multidimensional modulation. The amplitude, phase, and frequency-shift modulation are analog by integrating three-dimensional mechanical switching to a spiral slotted antenna. The maximum gain of the reconfigurable antenna can be adjusted in the type of 2 encoding bits at 9.5 GHz. Our work may pave the way to a high-performance reconfigurable antenna for 6G communication.
\end{abstract}
\smallskip

\section{Introduction}
Modulation and coding are crucial parts of a modern digital communication system\cite{proakis2001digital}, which has been widely studied. Cui et al. for the first time involved the groundbreaking concept of coding metamaterial and successfully achieve tunable  radiation patterns in the space-time domain\cite{cui2014coding}. This new class of signal interaction architecture may bring the communication system to a new era\cite{zhang2018space,ma2019smart, hodge2019reconfigurable,zhang2019breaking,liu2016convolution,wan2019multichannel,chen2019direct,wu2021reconfigurable}. Similarly, such a coding-based design concept might be potentially extended for intelligent microrobots\cite{ma2020programmable,zhang2020yin,han2020multi}.


Here, we propose an antenna based on spiral slotted waveguide featuring reconfigurable characteristics, both in the frequency domain and space domain. Both the polar angle and azimuth angle of radiation pattern varies when frequency and structural changes. The spatial distribution of the maximum gain of the designed structure was chosen to form 2 encoding bits referring to 4 encoding statuses, which can mimic a multidimensional modulation. Our work combines the modulation part and antenna part in the sending end, which paves the way to simplify the communication model.


\section{Design \& Theory}
A rectangular rotating plane featuring an area of \(22.86\times 10.16\) \si{mm^2} was designed to form structure by rotating along a spiral path. The rotating path can be written as a set of parametric equations. They are,

\begin{align}
X_t &= d\times t\times\cos\left(t\right)\\
Y_t &= d\times t \times\sin\left(t\right)\\
Z_t &= e\times t
\end{align}

\begin{figure}[ht]
\centering
\includegraphics[width=0.45\textwidth]{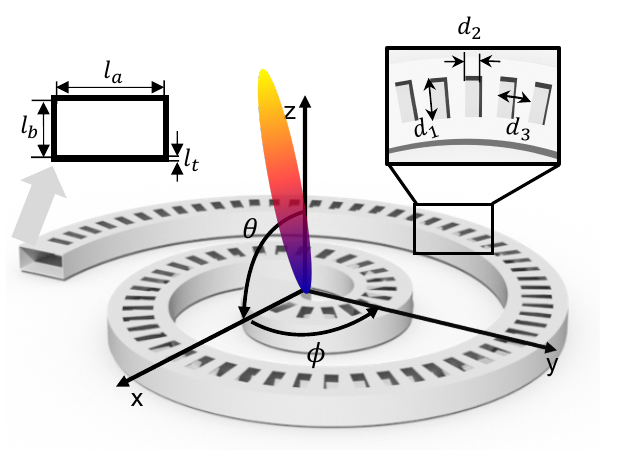}
\caption{The schematic of spiral slotted waveguide. \(l_a=22.86\)\si{mm}, \(l_b=11.06\)\si{mm}, \(l_t=1\)\si{mm}, \(d_1=10\)\si{mm}, \(d_2=5\)\si{mm} and \(d_3=10\)\si{mm}, the gradient filled ellipse represents the gain.}
\label{fig1}
\end{figure}

Here, \(t\) is the parameter adjusting total lengths of spiral wire and determine whether the path is clockwise or not. \(d\) denotes the variable responsible for changing the distance between each adjacent arms, and the vertical stretch degree \(e\) describes the height changes along \(z\) axis. As shown in Fig \ref{fig1}, we set \(t\) swept from \(-12\) to \(0\), \(d=10\), and \(e=0\). The rotating plane vertex was put to the origin of the path and made the whole plane perpendicular to the tangent of the spiral at the origin. Then the whole structure appeared while the rectangular shape rotating along the path with its short side keeping perpendicular to the \(x\)-\(y\) plane. After that, we hollowed the structure from both ends to a wall thickness of \(l_t\). At this stage, we got a rectangular pipe that can guide electromagnetic waves. For radiation, rectangular-shaped slots were drilled with the periodicity of \(d_3\) on the upper surface of the waveguide patterned by a period of \(d_3\) along the rotating path. Each slot was set to as \(d_1\) of length and width \(d_2\) of width.

There is a generally accepted reasoning saying radiation can be generated by slots because those narrow rectangular-shaped slots can be together equivalent to a phased dipole antenna according to the Babinet's principle\cite{born1999principles,balanis2016antenna,pozar2009microwave}. Suppose there is a charging particle moving along \(x\)-axis, the surface current density in the frequency domain induced by such a particle can be described as\cite{xi2009experimental}.

\begin{equation}
\vec{J}\left(\hat{r},w\right)=\hat{x}\frac{I\delta\left(\omega-\omega_0\right)}{2\pi\rho}\delta\left(\rho\right)e^{ik_xx}
\end{equation}

Where \(I\) is the current magnitude of each dipole, \(\rho\) represents the radial distance on the \(y\)-\(z\) plane, \(\delta\) means the Dirac delta function, \(\omega_0\) represents the working frequency, and \(k_x\) is the wave number of guided waves in rectangular waveguide. For a given frequency, the phase velocity of moving particles can be faster than the wave velocity of propagating waves in free space by taking \(v=\frac{\omega_0}{k_x}\)\cite{cheng1989field}. Radiation is generated from the waveguide upper surface to the free space in accordance with the related theory\cite{kong1975theory}. Furthermore, the radiation angle of the Cherenkov radiation is \(\theta_c=\cos^{-1}{\frac{c_m}{vn}}\), within which the \(c_m\) is the wave velocity in media, the \(v\) represents the particle moving speed, and \(n\) is the medium refractive index. Frequency can be modeled as a change of the moving speed of particles whereas variations of \(v\) will affect radiation angle \(\theta_c\), which also reflects a relation with azimuth, correspondingly. Moreover, stretching up structure from origin will lead to a hanging mosquito coil shape, and such a stretched shape will introduce azimuth changes, too. Thus, a radiation angle can be steered by changing working frequency in the spherical coordinate system. For our assumed spiral structure, the equivalent current moves along a spiral path, which will produce components in other directions. As a result, polar angle \(\phi\) changes due to constantly varying current components, while equivalent particles are moving along a spiral path. However, the variation range of the polar angle \(\phi\) in the operating frequency band is not large, usually in the range of \(90\)\si{\degree}. The problem to expand the reconfigurable range of the polar angle can be solved by rotating the entire structure according as C4 symmetry, which meets requirements of multidimensional modulation, in terms of frequency and space. Ultimately, four constellation points were observed on the observing plane, forming 2 encoding bits, adjusted by frequency and mechanical change.

\section{Simulation \& Result}

\begin{figure}[ht]
    \centering
    \begin{tabular}{@{}c@{} @{}c@{} @{}c@{}}
        \includegraphics[width=0.22\textwidth,keepaspectratio]{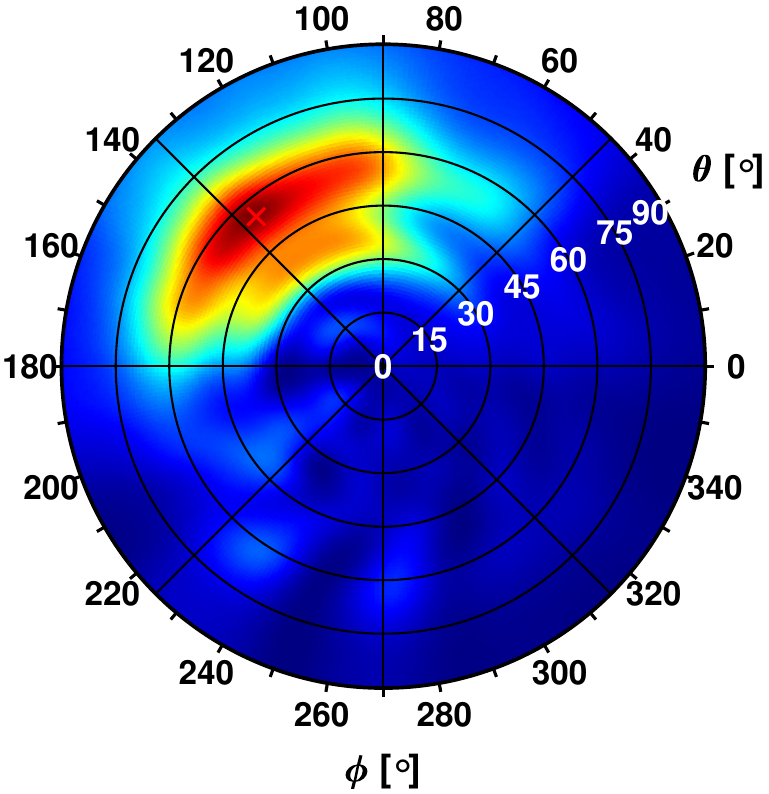}   & 
        \includegraphics[width=0.22\textwidth,keepaspectratio]{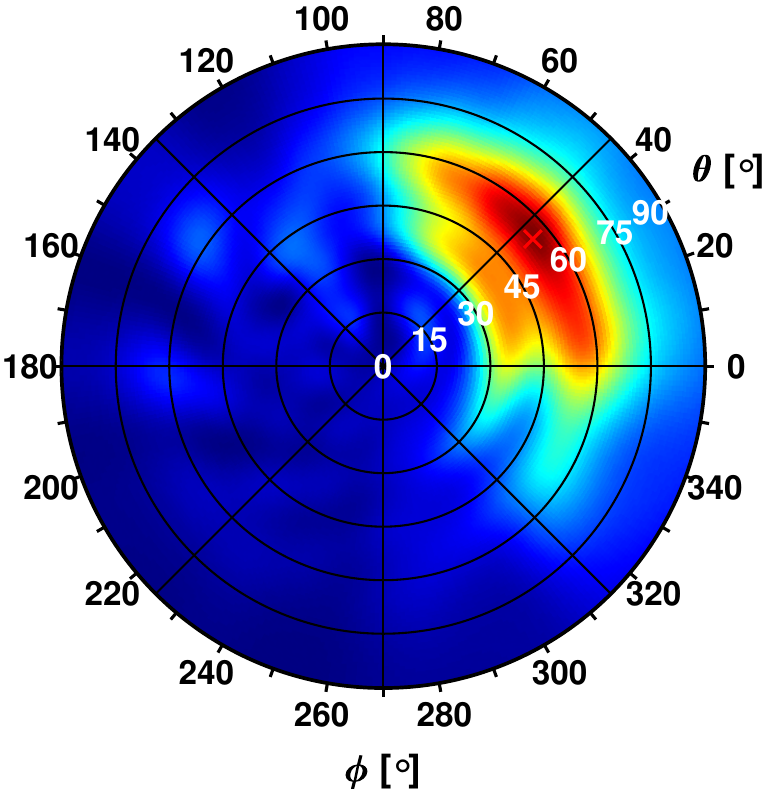}   &
        \multirow[t]{4}{*}[-0.146\textwidth]{\includegraphics[width=0.04\textwidth]{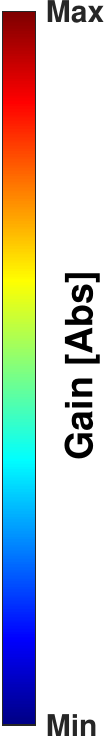}} \\
        (a) &   (c) &   \\
        \includegraphics[width=0.22\textwidth,keepaspectratio]{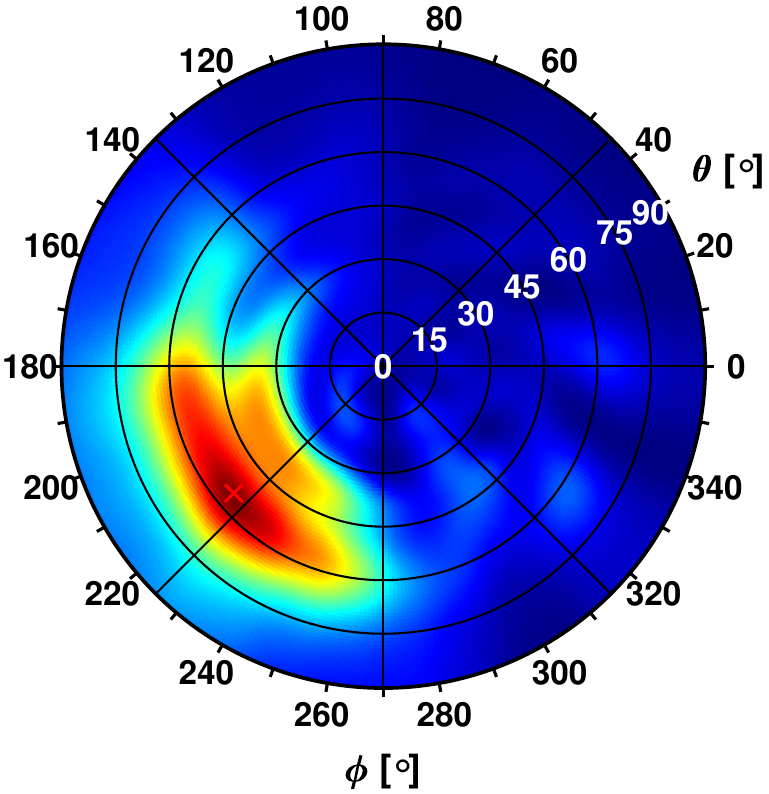}   & \includegraphics[width=0.22\textwidth,keepaspectratio]{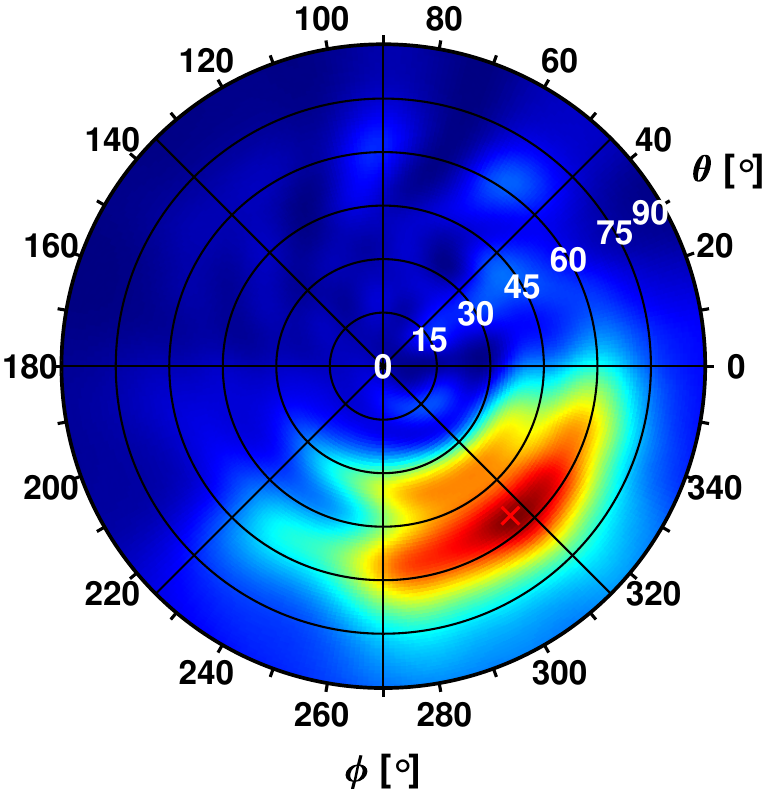}   &   \\    
        (b) &   (d) &   \\
    \end{tabular}
    \caption{Gain distributions at \(9.5\)\space\si{GHz} with structure rotating: a) 0\si{\degree}, b) 90\si{\degree}, c) 180\si{\degree}, and d) 270\si{\degree}. The maximum and minimum of gain are \(8.266\)\space\si{\dB} and \(0.0055\)\space\si{\dB}, respectively.}
    \label{fig2}
\end{figure}

A full-wave numerical software (CST microwave studio) was used to carry out the simulation. The frequency range was set from \SIrange{8}{14}{\GHz} with a step being \(0.5\)\space\si{\GHz}, and the observing plane was set to \(70\)\space\si{mm} above the ground. By taking frequency sweep and analyzing the gain distribution extracted from the far field monitor, for a non-rotation or stretch condition, the maximum gain occurred at \(131\)\si{\degree} of polar angle \(\phi\) and \(57\)\si{\degree} for azimuth angle \(\theta\) as shown in Fig \ref{fig2}a. 




\def\cdpw{0.45\columnwidth}
\def\cbpw{0.1\columnwidth}

Followed by the analysis of the relation of polar angle to frequency, an expected phenomenon can be observed that the location of maximum gain varies with frequencies. For the given frequency range, polar angle \(\phi\) tends to decrease with increasing frequency, on the whole. The \(\phi\) adjustable range falls to from \SIrange{65}{180}{\degree}, referring a \(115\)\si{\degree} of range, which is shown in Fig \ref{fig3}a, indicating that \(\phi\) decreases as frequency increases from \SIrange{8}{13.5}{\GHz}. And, it is easy to found that the azimuth increases by the upward stretch of structure. For example, the stretch distance was set to \(20\)\space\si{\mm} which refers to \(0.2\)\(\lambda\) for \(10\)\space\si{GHz} of the center frequency. At \(9.5\)\space\si{\GHz}, as shown in Fig \ref{fig3}b, we can see the azimuth changes only in a range of from \SIrange{35}{55}{\degree}. Then, by rotating the whole construct around \(z\)-axis counterclockwise by \(90\)\space\si{\degree}, \(180\)\si{\degree} and \(270\)\si{\degree} respectively, we can obverse that maximum gain has polar angle shifts with \(90\)\si{\degree}, \(180\)\si{\degree} and \(270\)\si{\degree} accordingly, making it appears in four different quadrants by rotation as shown in Fig \ref{fig2}a-\ref{fig2}d.

Moreover, by stretching up, the azimuth angle \(\theta\) increases comparing to a normal situation with about the \(16\)\si{\degree} of average, which refers to another possibility of spatial modulation. Finally, 4-status multidimensional modulation was realized within the observing plane, referring 2 encoding bits, leading to a frequency and spatial multidimensional modulation. The related constellation diagram is depicted in Fig \ref{fig3}c, which can be seen as strong evidence of functionalities of the proposed prototype.

\def\cdpwfthr{0.24\textwidth}
\begin{figure}[ht]
    \centering
    \begin{tabular}{c@{}c@{} c@{}c@{}}
        \includegraphics[width=\cdpwfthr]{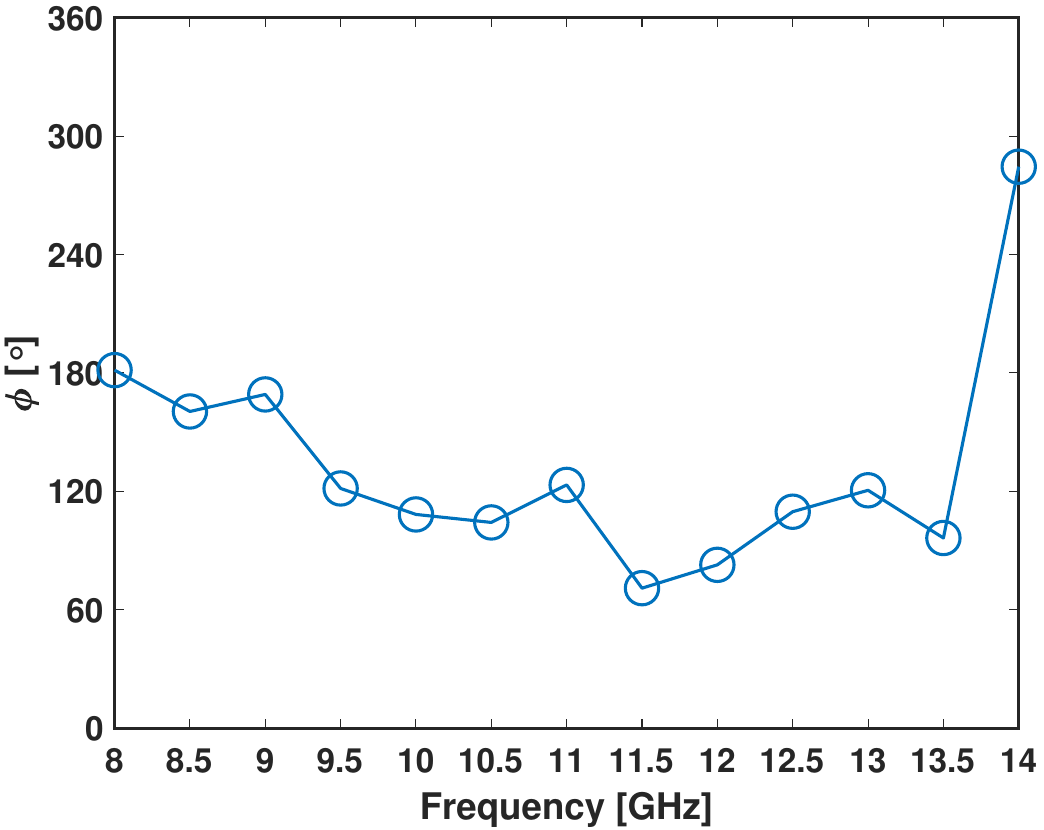} & \includegraphics[width=\cdpwfthr]{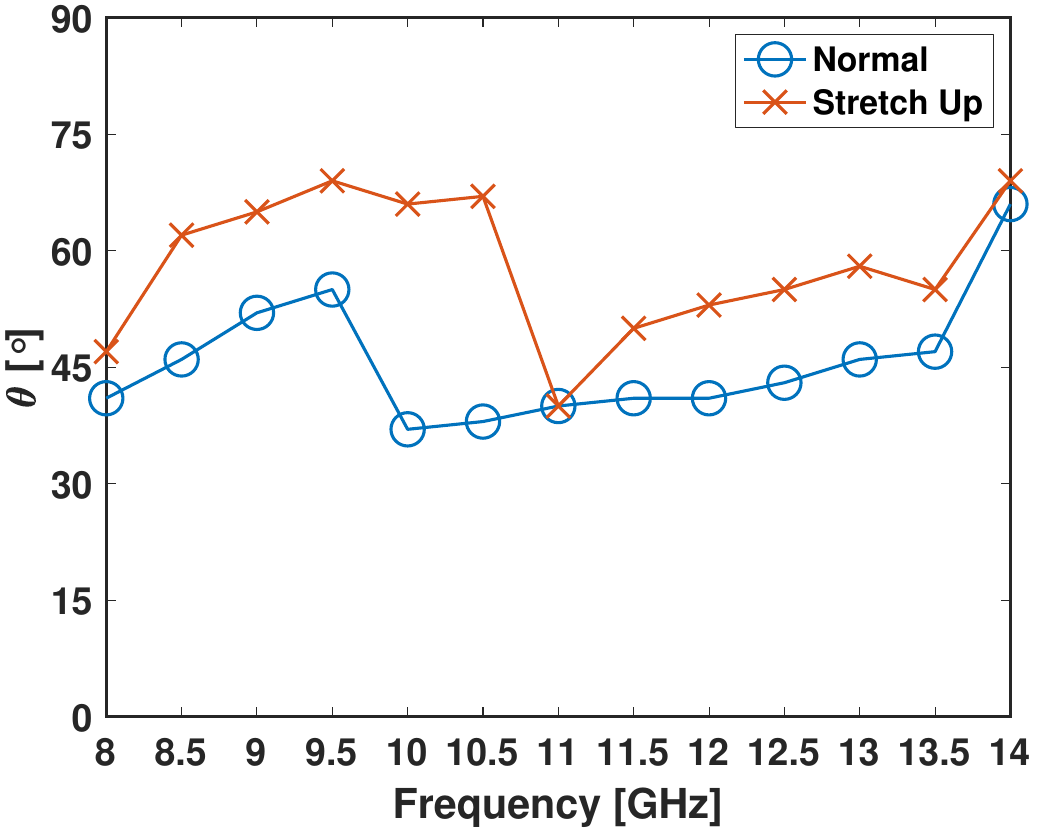} \\
        (a) &   (b) \\
        \multicolumn{2}{c}{\includegraphics[width=0.3\textwidth]{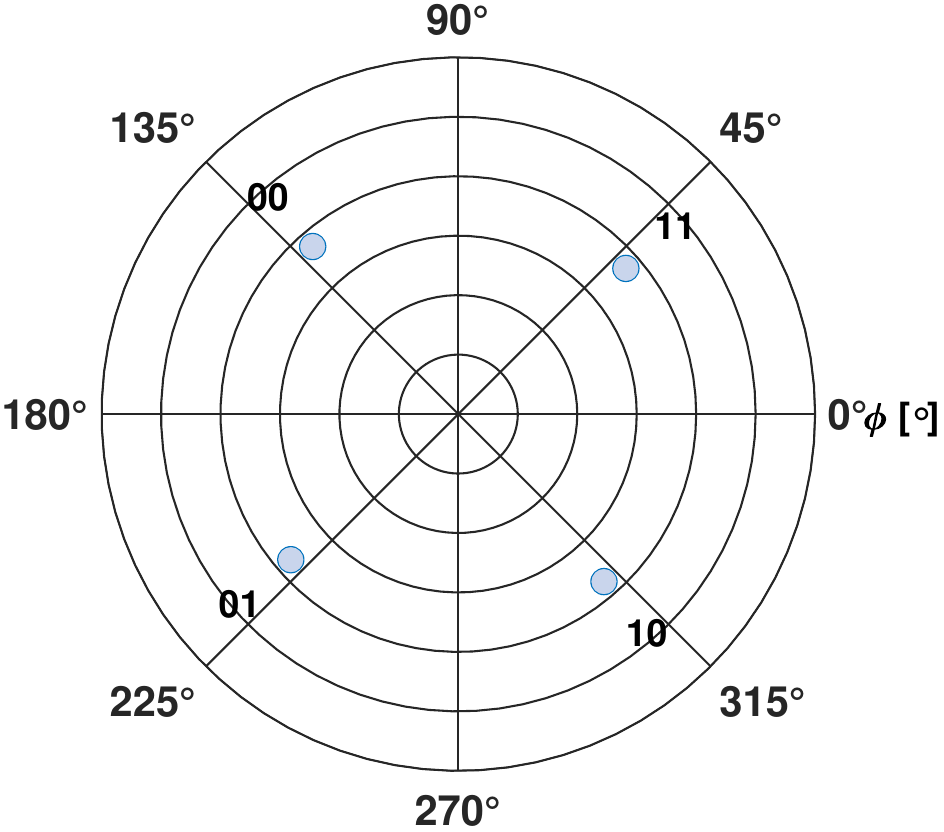}}   \\
        \multicolumn{2}{c}{(c)}
    \end{tabular}
\caption{Gain location and constellation diagram; \ref{fig3}a) polar angle \(\phi\) to frequency; \ref{fig3}b) azimuth angle \(\theta\) to frequency and stretch; \ref{fig3}c) 4-status constellation diagram.}
\label{fig3}
\end{figure}


\section{Conclusion}

In conclusion, we have proposed a spiral slotted waveguide antenna which can feature 2 encoding bits modulation at \(9.5\) \si{\GHz} by changing structure towards four different directions, which may pave the way to high-dimensional modulation catering requirements of both effectiveness and security for future wireless communication.


\bibliographystyle{IEEEtran}
\bibliography{references}
\end{document}